# Secondary spin current driven efficient THz spintronic emitters

*Piyush Agarwal[#], Yingshu Yang[#], Rohit Medwal[#], Hironori Asada, Yasuhiro Fukuma, Marco Battiato\*, and Ranjan Singh\**

P. Agarwal, Y. Yang, Prof. M. Battiato, Prof. R. Singh
Division of Physics and Applied Physics, School of Physical and Mathematical Sciences, Nanyang Technological University, 21 Nanyang Link, Singapore 637371, Singapore
E-mail: marco.battiato@ntu.edu.sg
E-mail: ranjans@ntu.edu.sg

P. Agarwal, Prof. R. Singh
Center for Disruptive Photonic Technologies, The Photonics Institute, Nanyang Technological University, Singapore 639798, Singapore
E-mail: ranjans@ntu.edu.sg

Dr. R. Medwal
Department of Physics, Indian Institute of Technology Kanpur, Uttar Pradesh 208016, India

Prof. H. Asada
Department of Electronic Devices and Engineering, Graduate School of Science and Engineering, Yamaguchi University, Ube 755-8611, Japan

Prof. Y. Fukuma
Department of Physics and Information Technology, Faculty of Computer Science and System Engineering, Kyushu Institute of Technology, Iizuka 820-8502, Japan

[#]Equal Contribution

**Abstract**

Femtosecond laser-induced photoexcitation of ferromagnet (FM)/heavy metal (HM) heterostructures have attracted attention by emitting broadband terahertz frequencies. The phenomenon relies on the formation of ultrafast spin current, which is largely attributed to the direct photoexcitation of the FM layer. However, we reveal that during the process, the FM layer also experiences a secondary excitation led by the hot electrons from the HM layer that travel across the FM/HM interface and transfer additional energy in the FM. Thus, the generated secondary spins enhance the total spin current formation and lead to amplified spintronic terahertz emission. The results also emphasize the significance of the secondary spin current, which even exceeds the primary spin currents when FM/HM heterostructures with thicker HM are used. An analytical model is developed to provide deeper insights into the microscopic processes within the individual layers, underlining the generalized ultrafast superdiffusive spin-transport mechanism.

# 1. Introduction

Out-of-equilibrium excitation of spintronic materials using femtosecond laser pulses [1] has drawn immense interest in condensed matter physics by exploiting asymmetricity in the lifetime and velocity of excited majority and minority spins [2,3]. Upon laser absorption, the FM/HM heterostructure undergoes spin excitation, giving rise to spin flipping and scattering events in femtosecond timescale [4–7]. At the same time, the spins superdiffuse semi-ballistically, forming direct channels of spin-transport between the FM and HM layers [8–11]. The phenomena has led to the development of a range of applications such as broadband terahertz sources [12–15], ultrafast spin-transfer-torque driven logic devices [16], terahertz (THz) magnetometry [17–19], spin-resolved electron spectroscopy [20], ballistic electron emission microscopy [21], next-generation data-processors [22–26], and ultrafast spin injection in semiconductors and topological insulators [27–31]. Theoretical calculations attributed to the superdiffusion predicted the scattering processes within the FM and HM to cause the flow of hot electrons both from the FM to HM and HM to FM [8,32], leading to enhanced ultrafast demagnetization of FM [32,33]. However, the underlying processes of spintronic terahertz emission have so far discussed only the primary photoexcitation of the FM layer, assuming unidirectional carrier flow from FM to HM. Instead, our results, in agreement with the theoretical predictions [8,32] demonstrate that a simultaneous photoexcitation of the HM layer induces additional carrier flow from HM to FM and plays a crucial role in the generation of enhanced terahertz radiation.

The photoexcited carriers in the HM layer are spin-unpolarized and do not possess energy-dependent spin asymmetry. Yet, the excited hot electrons experience a spin-differentiated diffusion as they travel across the FM/HM interface and deposit additional energy in the FM layer, leading to the secondary excitation of the FM layer[8,9,32,33]. Hence, the FM generates an additional spin-polarized current that constructively enhances the produced spin current [8,9,33,34]. In this work, we quantify the effect of secondary FM excitation by investigating the terahertz radiation from the FM/HM heterostructure with varying thicknesses of the HM layer from 0 to 10 nm. Upon increasing the HM thickness, three correlated phenomena arise, (i) increase in absorption of terahertz radiation within the HM layer, (ii) increase in spin-to-charge conversion, and (iii) increase in the secondary spin current. We decouple the three phenomena and realize strong evidence for an enhanced carrier generation in the FM layer. The theoretical model agrees well with the experimental results and unveils the microscopic processes that contribute to superdiffusion. Moreover, the secondary spin currents even surpass the primary spin currents when heterostructures with higher HM thickness are used. The results thus emphasize the significance of primary and secondary excitation for ultrafast spin transport processes and establish a generalized spin transport mechanism crucial for developing spintronics-based THz functional devices.

## 2. Results and Discussions

**Figure 1(a)** illustrates the spin-flip scattering and spin-transport processes [35] that constitute the ultrafast demagnetization in the spintronic heterostructure. So far, the spin-flip scattering (process 1) and the spin transport (process 2) have emphasized only the primary photoexcitation of the FM layer. However, as highlighted in process 3, the HM layer also experiences photoexcitation, due to which HM excites the FM once again, leading to an enhanced superdiffusive spin current, $j_s$. In principle, when the spins from FM superdiffuse into the HM layer, they undergo an inverse spin Hall effect (ISHE) and generate a transient ultrafast charge current, $j_c$, where $j_c = \gamma j_s \times M/|M|$ [14], $\gamma$ is the spin Hall angle of the HM layer, and $M$ is the magnetization vector of the FM layer. As a result, the heterostructure emits terahertz electric field according to $E(t) = \partial j_c(t)/\partial t$ [12–14,36] and terahertz pulse amplitude scales with the ultrafast charge current amplitude, given by, $E_{(t)}^{peak} \propto j_c^{peak} \propto j_s^{peak}$ [17,37]. Besides, within such a heterostructure, the terahertz field is also produced from minor processes that do not involve superdiffusion of spin current across the FM/HM interface. The mechanism includes 1) ultrafast change in magnetic dipoles [19], 2) ultrafast change in electric dipoles [19], and 3) back reflection of spins in FM that experience ISHE within itself or Subs/FM interface [38,39]. In Figure 1(b and c), a comparison of the terahertz generation is performed between two spintronic heterostructures, Pt(0nm)/$Ni_{80}Fe_{20}$(3nm) [PT-0] and Pt(2 nm)/ $Ni_{80}Fe_{20}$ (3nm) [PT-2]. From this point, $Ni_{80}Fe_{20}$ (permalloy) is written as NiFe. Refer to Supplementary Sections S1, S3, S4, and S5 for setup and experimental methods. The sample without heavy metal, [PT-0], is chosen to inhibit the superdiffusive spin-transport and represents the THz generation arising only due to the minor processes. On the other hand, [PT-2] exhibits superdiffusive spin transport along with minor processes. The resulting terahertz electric field emission is shown in Figures 1(d and e), where the terahertz pulse amplitude inclusive of superdiffusion (Figure 1(e)) is observed to be an order of magnitude higher than the counterpart in Figure 1(d). As such, the terahertz pulse amplitude essentially provides a measure of superdiffusive spin current and has a much lower contribution from the minor processes. The contributions from the minor processes are later separately accounted for in the theoretical analysis.

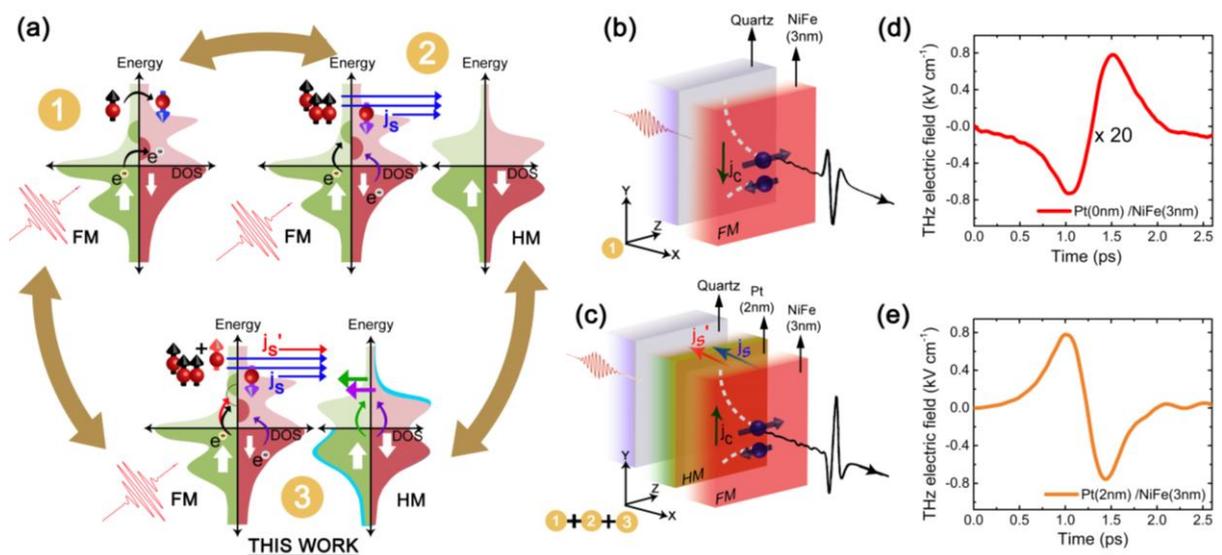

**Figure 1:** (a) Illustration of three major ultrafast demagnetization channels upon femtosecond laser pulse excitation of the spintronic heterostructure. Process 1 and 2: spin-flip mechanism

and spin-transport mechanism, respectively, led by the primary photoexcitation of FM. Process 3: [This work] spin-transport mechanism highlights FM's secondary excitation, initiated by hot carriers from the photoexcited HM layer. (b) Schematic of Quartz/NiFe heterostructure without HM layer to inhibit the superdiffusion process. (c) Schematic of Quartz/Pt/NiFe heterostructure exhibits terahertz emission propelled by both primary and secondary excitations alongside the minor processes. (d) Measured terahertz electric field emitted from photoexcited Quartz/NiFe. (e) Measured terahertz electric field emitted from photoexcited Quartz/Pt/NiFe.

Further, we consider the net spin current generated in heterostructure to consist of two constituent components represented by $j_s$ and $j_s'$. Here, the primary spin current, $j_s$ arise from the direct photoexcitation of the FM layer, and secondary spin current, $j_s'$ arise due to HM-led photoexcitation of the FM layer (**Figure 2(a)**). To decouple the two components experimentally, we used a set of samples as Quartz(1mm)/ Pt($d$ nm)/ NiFe(3 nm) with $d$ varying from 0nm → 10nm and is referred as [PT-$d$]. With the increase in HM layer thicknesses, three coupled effects arise: (1) increased absorption of terahertz radiation within the HM layer, (2) increase in spin-to-charge conversion, (3) enhancement of $j_s'$. The transmission from all the samples are recorded to estimate the attenuation of terahertz due to absorption in the HM layer. A separate ZnTe-based time-domain terahertz spectrometer was used to perform the transmission measurement, with spintronic emitters at the terahertz focus point (refer to Supplementary Section S1 for experimental details). In Figure 2(b), we can observe an exponential decrease in terahertz transmission amplitude with increasing HM layer thicknesses. As a control experiment, the transmission was also recorded in the photoexcited state of the heterostructure, where an additional laser beam ($\lambda$ = 800 nm, 35 fs, 1 kHz repetition rate) was used to photo-illuminate the sample (see inset of Figure 2(b)). The terahertz transmission was found to be almost identical in both photoexcited and non-excited states of the sample. Further, to estimate the increase in spin-to-charge conversion and the $j_s'$, the terahertz emission from all the samples were recorded as shown in Figure 2(c). The emission initially increases up to [PT-2] in the left region but exhibits an exponential decrease beyond [PT-4]. The region in the right (where $d \geq$ 4nm) is of particular interest where the HM thickness is more than the spin diffusion length of Platinum, $\lambda_{sd}$=1.3nm [40,41].

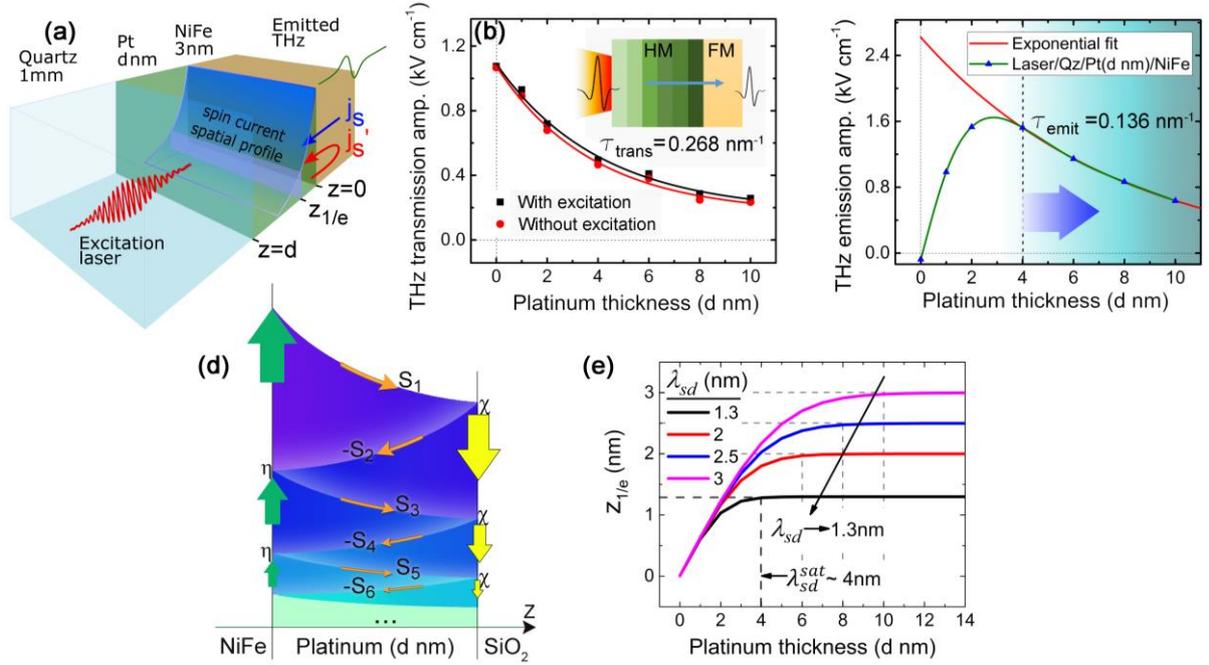

**Figure 2:** (a) Illustration of the superdiffusion process due to the photoexcitation of both FM and HM layers. Spin current, $j_s$ is led by the primary photoexcitation, and $j_s'$ is led by the secondary excitation of the FM layer. The spatial profile indicates the decay of spin polarization in the HM layer. $z_{1/e}$ marks the depth in the HM layer where the spin polarization decreases to $1/e$ of its strength at $z=0$ (b) Experimental observation of terahertz transmission decay as observed in [PT-$d$] where $d$=0,1,2,4,6,8,10 nm. Identical decay was observed at both photoexcited (as seen in the inset) and non-photoexcited states; $\tau_{trans} = 0.268 \ nm^{-1}$. (c) The peak-to-peak amplitude of terahertz emission from all the samples. The dashed line divides the region, indicating the exponential decrease in [PT-$d$] when $d \geq$ 4nm. A solid red line indicates an exponential fit for emission in [PT-$d$] when $d \geq$ 4nm; with $\tau_{emit} = 0.136 \ nm^{-1}$. (d) The construction of spatial profile describes the decay of polarized spins through multiple spin

reflections at the surface with the *n*th spin decay term given by $\vec{S_n}(z)$. (e) The formalism suggests that $z_{1/e}$ saturates at different depths of HM layer ($\lambda_{sd}^{sat}$) with varying $\lambda_{sd}$. In our case of Platinum, $\lambda_{sd}$=1.3nm yields $\lambda_{sd}^{sat}$ ~ 4nm.

A monotonic exponential decrease beyond [PT-4] suggests a saturated spin-to-charge conversion; however, in contrast, one would expect the exponential decay to appear beyond $\lambda_{sd}$=1.3nm. To understand the behavior, we develop a formalism and model the spin-to-charge conversion through estimating the spatial decay of polarized spins in the HM layer (Figure 2(d)). Within such nanometer-thin films of HM, the spin reflection at the surfaces becomes significant, forming the left- and right-propagation [42]. Thus, by applying the continuity boundary conditions of the fields at the interfaces, the propagation of the spins in HM is described using a generalized Transfer Matrix Method (TMM). Moreover, for reproducing the emission process, we applied a modified TMM [42] that treat HM as a layer for the THz source. The modified TMM yields a volume charge current, $\boldsymbol{j_c}$ [*t*,*z*] with a temporal and spatial dependence parallel to the sample surface. The temporal profile of $\boldsymbol{j_c}$ is used to describe the transient THz radiation field, calculated in accordance to the experimental shape of the THz emission profile. The spatial profile thus accounts for the total $\boldsymbol{j_s} \rightarrow \boldsymbol{j_c}$ along the HM layer, as shown in Figure 2(d). Here, the *n*th spin decay term $\vec{S_n}(z)$ exhibit exponential decay as given by equations below, but the overall spin dynamics profile becomes more complicated than a simple exponential decay [43] due to both left and right direction propagation.

$$\vec{S_1}(z) = e^{-\frac{z}{\lambda_{sd}}} \; ; \; \overleftarrow{S_2}(z) = \chi e^{-\frac{d}{\lambda_{sd}}} * e^{-\frac{d-z}{\lambda_{sd}}}$$

$$\vec{S_3}(z) = \chi\eta e^{-\frac{2d}{\lambda_{sd}}} * e^{-\frac{z}{\lambda_{sd}}} \; ; \; \overleftarrow{S_4}(z) = \chi^2\eta e^{-\frac{3d}{\lambda_{sd}}} * e^{-\frac{d-z}{\lambda_{sd}}}$$

where, *z* is the axis along the thickness of the HM layer, $\chi$ is the spin reflection at the Pt/Quartz interface, $\eta$ is the spin reflection at NiFe/Pt interface, $\lambda_{sd}$ is the spin diffusion length, and *d* is the thickness of the HM layer. The spin losses due to reflections at the interface are almost negligible, and we assume $\chi = \eta = 1$, which is in agreement with the previous studies [14]. The direction of an arrow depicts the spin decay path in $S(z)$. As given in Figure 2(d), the overall spatial profile $S(z)_{overall}$ can therefore be constructed as,

$$S(z)_{overall} = (\overrightarrow{S_1} + \overrightarrow{S_3} + \overrightarrow{S_5} + \cdots) - (\overleftarrow{S_2} + \overleftarrow{S_4} + \overleftarrow{S_6} + \cdots)$$

$$S(z)_{overall} = \frac{e^{-\frac{z}{\lambda_{sd}}} - \chi e^{-\frac{2d}{\lambda_{sd}}} * e^{\frac{z}{\lambda_{sd}}}}{1 - \chi\eta e^{-\frac{2d}{\lambda_{sd}}}} \quad [1]$$

We use the formalism to estimate the HM thickness required for a spin-to-charge conversion. The maximum amplitude of the spins near the FM/HM interface is calculated using Equation 1. For example, at $z=0$, $S(z) = S(z_0) = \frac{1-\chi e^{-\frac{2d}{\lambda_{sd}}}}{1-\chi\eta e^{-\frac{2d}{\lambda_{sd}}}}$ and $S(z)$ reaches $S(z_0)/e$ in the depth of the HM layer, given by $z = z_{1/e}$. Using Equation 1, $z_{1/e}$ can thus be calculated as

$$z_{1/e} = \lambda_{sd} Log_e \left[\frac{1}{2}e^{-1+\frac{2d}{\lambda_{sd}}}\left(-1 + e^{-\frac{2d}{\lambda_{sd}}} + e^{-\frac{2d}{\lambda_{sd}}}\sqrt{e^{\frac{4d}{\lambda_{sd}}}\left(1 + 4e^{2-\frac{2d}{\lambda_{sd}}} + e^{-\frac{4d}{\lambda_{sd}}} - 2e^{-\frac{2d}{\lambda_{sd}}}\right)}\right)\right] \quad [2]$$

Here, at the known spin diffusion length for Platinum, where $\lambda_{sd}$ =1.3nm [40,41], $z_{1/e}$ is observed to saturate at Platinum thickness of ~4nm (Figure 2(e)). In other words, the Platinum layer with $\lambda_{sd}^{sat}$~4nm can be interpreted as the thickness required for saturating spin to charge conversion with spin diffusion length $\lambda_{sd}$ =1.3nm. Thus, no additional spin-to-charge conversion occurs in HM when its thickness is increased beyond 4nm. Recalling from Figure 2(c), the result holds exactly true, and therefore a monotonic exponential decay was observed beyond [PT-4]. With the given implementation, we simplify the visualization of THz emission as illustrated in Figure 3(a). Since we consider only the ultrafast spin excitations (spin-flip and spin scattering processes) occurring in sub-picosecond timescales, the terahertz emission resulting from the $j_s'$ and $j_s$ can be assumed to interfere constructively. Here, the terahertz emission from HM/FM sample was detected from the FM side, as shown in Figure 3(a). Such a scenario consists of two main phenomena i) the superdiffusive spin current experiencing ISHE in the HM layer generates the terahertz radiation in both parallel and anti-parallel directions of laser propagation (indicated as paths A and B), and ii) the emission in path B experiences a back-reflection which constructively interfere with the emission from path A. Note that, for samples [PT-$d$], with $d \geq 4$ nm, the relative terahertz radiation, $\frac{\text{terahertz amplitude}}{\text{fluence absorbed}}$, emitted in the direction of path A must remain unchanged with increasing Platinum thickness. However, the emission from back-reflected path B would experience an additional terahertz absorption depending on the thickness of respective HM layer. Therefore, the total emitted terahertz radiation from the FM/HM heterostructure must have an identical decay rate as the terahertz transmission decay performed using terahertz time-domain spectroscopy. Besides, the terahertz emission decay can still deviate with respect to the transmission decay if the contribution from the $j_s'$ is introduced, which is dependent on the HM layer thickness. Figure 3(b) highlights the three probable outcomes after comparing the terahertz emission with the terahertz transmission decay that can provide both qualitative and quantitative evidence of the presence of $j_s'$.

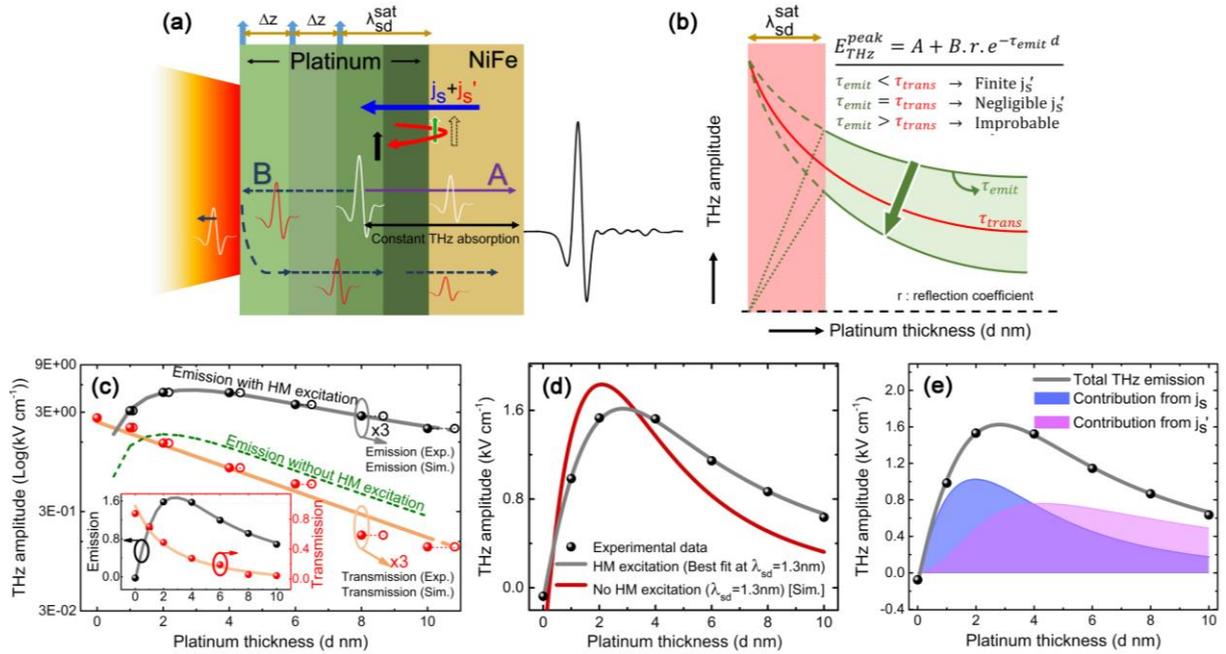

**Figure 3:** (a) Experimental illustration to visualize the $j_s'$ dependence on the HM layer's thickness. With increasing HM thickness, the terahertz emission follows three correlated effects (1) increased absorption of terahertz radiation within the HM layer, (2) increase in spin-to-charge conversion, (3) enhancement of $j_s'$. (b) For heterostructures with Pt thickness more than $\lambda_{sd}^{sat}$, $E_{THz}^{peak} = A + B.r.e^{-\tau_{emit}d}$, where A and B are the peak amplitude of emission arising from paths A and B, as shown in 3(a). $\tau_{emit}, \tau_{trans}$ are the decay rate of terahertz emission and terahertz transmission, respectively. $r$ is the reflection coefficient, and $d$ is the thickness of HM. The schematic highlights the relation between the decay rate of transmitted terahertz radiation (solid red curve) and the decay rate of emitted terahertz radiation (solid green curve). (c,d,e) Black spheres show the experimentally observed THz emission. The solid black line shows the theoretical model's agreement with the experimental THz emission accounting for both primary and secondary excitation, as defined in Equation 4. (c) Red spheres show the experimentally observed THz transmission modeled with exponential decay, as shown by a solid red line. Corresponding to each red and black sphere, a thick circle is connected to show the extent of error introduced due to the 8.2% error in Platinum thickness. The error bar in THz amplitude is omitted due to a low ~0.94% fluctuation. Note that even the inclusion of the error bar produces identical transmission and emission decay rates. Refer to Supplementary Section S4 for error analysis. The log scale data provides a direct comparison of the decay rate observed through the slope. The inset shows the THz transmission and THz emission on a linear scale. (d) A solid red line exhibits disagreement of the theoretical model when secondary excitation is not considered. (e) The blue and pink region demonstrates the individual contribution of the $j_s$ and $j_s'$ in the total terahertz emission, respectively.

Clearly, a slower decay rate of terahertz emission ($\tau_{emit} = 0.136\ nm^{-1}$) with respect to the transmission ($\tau_{trans} = 0.268\ nm^{-1}$), can be observed in Figure 3(c), indicating that the increasing HM layer must be creating an additional increase in the spin current. The significant difference in slope is the direct experimental evidence of the presence of $j_s{'}$, which occurs due to the HM layer. To visualize the effect theoretically, the THz transmission profile is modeled using the transfer matrix method (TMM) [42,44] by considering the permittivities of the films at terahertz frequency. As such, the standard terahertz emission (governed by $j_s$) from the primary photoexcitation of NiFe is given in Equation 3 as.

$$E_{THz}(d) = \alpha E_{THz}^{Pt}[\lambda_{sd}, d] A_{excit}^{NiFe}[d] + \xi E_{THz}^{NiFe}[\lambda_{NiFe}, d] A_{excit}^{NiFe}[d] \quad [3]$$

The first term accounts for the contribution to terahertz radiation from the FM → HM superdiffusion process (primary excitation), and the second term accounts for the contribution from the minor process that does not include the role of Pt, as discussed in Figure 1(b and c). $\alpha$ and $\xi$ are scaling parameters, respectively. $E_{THz}^{Pt}[\lambda_{sd}, d]$ and $E_{THz}^{NiFe}[\lambda_{NiFe}, d]$ refers to the THz emission undergoing spin-to-charge conversion in Pt and NiFe, respectively. The transient THz electric field and ultrafast spin dynamics profile are used to calculate the emission matrix element. The calculation incorporates the total absorption of excitation laser in each individual layer of the heterostructure given by $A_{excit}^{NiFe}[d]$. Refer to the Supplementary section for experimentally measured transmission and reflection percentage from the heterostructure where the Poynting's theorem was implemented to find the fluence absorption in the individual layers of FM and HM. Nevertheless, the equation 3 with all possible characteristic parameters still fails to explain the HM thickness-dependent terahertz emission profile (refer to the solid red line in Figure 3(d). Therefore, a modified model, including the HM-led secondary excitation of FM, is used, as given in Equation 4.

$$\begin{aligned} E_{THz}(d) = &\ \alpha E_{THz}^{Pt}[\lambda_{sd}, d] A_{excit}^{NiFe}[d] + \xi E_{THz}^{NiFe}[\lambda_{NiFe}, d] A_{excit}^{NiFe}[d] \\ &+ \beta E_{THz}^{Pt}[\lambda_{sd}, d] A_{excit}^{Pt}[d] \int_0^d e^{-\frac{z}{\lambda_E}} dz \end{aligned} \quad [4]$$

Note that the third correction term indicates the aforementioned $j_s'$, with $\beta$ as the scaling factor. The term incorporates the absorption of the excitation laser in Pt, $A_{excit}^{Pt}[d]$ and $\int_0^d e^{-\frac{z}{\lambda_E}} dz$ signifies the total amount of energy that Pt diffuses into the NiFe with $\lambda_E$ representing the energy diffusion length. Using Equation 4, we achieve an excellent correspondence between the theoretical THz emission profile (solid black line in Figure 3(d)) and the experimental THz emission profile (black spheres in Figure 3(d)), where we find the best fit at $\lambda_{sd} = 1.3$ nm [14] with $\lambda_E = 0.74$ nm (Figure 3(c-e)). In addition, if we calculate back the THz emission with only the first two terms (considering $j_s'=0$) keeping $\lambda_{sd} = 1.3$ nm, the overall decay rate of the emission (dashed green line in Figure 3(c)) exhibit an identical decay as that of transmission (red spheres in Figure 3(c)). The observation is in complete agreement with the relative decay behavior of the terahertz transmission and terahertz emission as a function of Platinum thickness, shown in Figure 3(b). Further, upon decoupling the contribution of the primary spin current, $j_s$ and the secondary spin current, $j_s'$, a significant role of HM layer driven $j_s'$ is explicitly seen in Figure 3(e).

In summary, we highlight that although the HM layer generates unpolarized spins upon femtosecond laser excitation, the HM forms a spin-differentiated diffusion across the FM/HM interface and deposits additional energy to the FM layer. As a result, the FM undergoes a secondary spin excitation apart from the primary photoexcitation. The experimental and theoretical results indicate a prominent contribution of the secondary spin current generation in comparison to the primary spin current and provide a measure of energy transfer from HM to FM at an ultrafast time scale. As such, we unveil the contribution of primary and secondary spin current driven superdiffusion in FM/HM heterostructure that will be essential for developing efficient terahertz sources and ultrafast spintronics for future technologies.

**Experimental and theoretical method details are** given in the Supplementary Information**,** linked to the online version of the paper**.**

**Acknowledgements**

R.S. and P.A. would like to acknowledge National Research Foundation, Singapore, for the support through NRF-CRP23-2019-0005. M.B. would like to acknowledge Nanyang Technological University, NAP-SUG grant.

**Author Contributions**

P.A., R.M., and R.S. conceived the project and designed the experiments. P. A. performed all the measurements and experimental analysis. Y.Y. and M.B. provided the theoretical model. H.A., Y.F. fabricated the spintronic emitter. All the authors analyzed and discussed the results. P.A., Y.Y., R.M., and R.S. wrote the manuscript with inputs from all the authors. R.S. lead the overall project.